\documentclass[preprint,showpacs,preprintnumbers,amsmath,amssymb]{revtex4}
                           
\usepackage{graphicx}
\usepackage{dcolumn}
\usepackage{bm}

\begin{document}


\title{Numerical Study of the Cahn-Hilliard Equation in One, Two and Three Dimensions}

\author{E. V. L. de Mello}
\affiliation{%
Instituto de F\'{\i}sica, Universidade Federal Fluminense, Niter\'oi, RJ 24210-340, Brazil\\}%

\author{Otton Teixeira da Silveira Filho }
\affiliation{%
Instituto de Computa\c{c}\~ao, Universidade Federal Fluminense, Niter\'oi, RJ 24210-340, Brazil\\}%

\date{\today}

\begin{abstract}

The Cahn-Hilliard equation is related with a number of interesting
physical phenomena like the spinodal decomposition, phase separation
and phase ordering dynamics.
On the other hand this equation is very stiff an the difficulty to solve it 
numerically  increases with the dimensionality 
and  therefore, there are several published numerical studies  in one dimension 
(1D), dealing with different approaches, and much fewer
in two dimensions  (2D). In three dimensions (3D) there are very few publications, 
usually concentrate in some specific result without the
details of the used numerical scheme.
We present here a  stable and fast conservative finite difference scheme
to solve the Cahn-Hilliard with two improvements: a splitting potential into a
implicit and explicit in time part and a the use of free boundary 
conditions. We show that gradient stability is achieved 
in one, two and three dimensions with large time marching
steps than normal methods.
\end{abstract}

\maketitle

\section{Introduction}

The theory of the phase-ordering dynamics following a rapid cooling
down (or quenching) through the critical temperature from a homogeneous or disordered
phase into an inhomogeneous or ordered state has been studied for
decades\cite{Lifshitz,HH,Gunton,Bray}. This phenomenon is known as
spinodal decomposition. Part of the fascination of the
field is because the ordering does not occur instantaneously. Instead,
a network of domains of the equilibrium phases develops, and the
typical length scale associated with these domains increases with
the time. In order words, the length scale of the ordered regions
growth with time as the different (broken-symmetry) phases compete
in order to achieve the equilibrium state. One of the leading
models devised for the theoretical study of this phenomenon is 
based on the Cahn-Hilliard formulation\cite{CH}. The Cahn-Hilliard
(CH) theory was originally proposed to model the quenching of binary
alloys through the critical temperature\cite{CH} but it has subsequently 
been adopted to model many other physical systems which go through a 
similar phase separation\cite{HH,Gunton,Bray}. 

Recently it has been discovered that high critical temperature
superconductors have an intrinsic inhomogeneous phase which
exhibit patternings which involve nanoscale regions of phase separations,
often referred as stripes as revealed by neutron diffraction
studies\cite{Tranquada} and EXAFS\cite{Bianconi}. 
Similar findings were measured by
very fine scanning tunneling microscopy/spectroscopy
(STM/S) data\cite{Fournier,Pan} which have shown
the local charge and the superconducting gap spatial 
variation through the  differential conductance.
The ubiquity of such
intrinsic inhomogeneities, as well as the importance of the materials 
that exhibit them has motivated intense experimental and theoretical
research into the details of the phenomena\cite{OWK,Mello1,Mihailovic,Mello2,Ictp}.
Recently a theory of the critical field $H_{c2}$ 
based on a distribution of superconducting regions
with different critical temperatures due to these intrinsic inhomogeneities
has explained some nonconventional features of current data on 
some cuprates\cite{Edson}.
The  manganites which exhibit a colossal magnetoresistance have also some
properties which may be linked with clusters formation upon cooling or an 
intrinsic phase separation\cite{Dagotto,Dagotto1}. Researches on these
materials are currently been investigated by a sizable fraction
the condensed matter community. Therefore, studies
based on the CH equation may be
useful to understand the puzzle of the origins of 
such intrinsic phase separation in these materials and how it
affects their physical properties. Here we want to study specifically
the Cahn-Hilliard equation and we will deal with the problem of
phase separation in high-$T_c$ superconductors in another publication.

In an alloy system composed by a binary mixture we can define
the local phase variable $u(\vec x)$ as the difference in concentration between the
two (incompressible) components or simply, the concentration of one of 
the components over the domain $\Omega$ ($\vec x \in \Omega$) which represents
the system. It is clear that this type of phase variable or
order parameter is always
conserved for an isolated system and we will use below this property as 
the main guide for our 
numerical method. In the CH theory, the time variation of the
order parameter $u(\vec x,t)$ is given in terms of the functional
derivative of a time-dependent free-energy functional $F(u)$ leading
to an equation of motion to diffusive transport of the order 
parameter, namely
\begin{eqnarray}
\frac{\partial u}{\partial t} = M\nabla^2\frac{\delta f}{\delta u}
\label{Eq1}
\end{eqnarray}
where $f$ is the free-energy density and the (constant) mobility $M$ will hereafter
be absorbed into the time scale although there are cases in which
the mobility can be a function of the position\cite{Kim}. The free-energy functional is 
assumed, to most of the physical applications, to follow the
Ginzburg-Landau (GL) form:
\begin{eqnarray}
f= {{{1\over2}\varepsilon^2 |\nabla u|^2 +V(u)}} 
\label{Eq2}
\end{eqnarray}
where the potential $V(u)$ may have a double-well structure like, for
instance, $V(u)=(u^2-1)^2/4$  as many authors have 
used\cite{Eyre,Furihata,Elliot1,VLR} and, for the sake of 
comparison  with these previous works, we will adopt it 
below. There are other possibilities like $V(u)= au^2/2+bu^4/4+...$
which is more convenient for  physical applications since,
usually the GL free energy is a power 
expansion in $u$ with coefficients ($a$, $b$, ...) which depend on
the temperature,
the applied field, and on other physical properties. 
It is easy to see that the above double well potential will favor
two phases with densities $u=\pm 1$. If one uses a potential with
three minima, it will appear three major phases and so on. Bray\cite{Bray}
pointed out that one can explore the fact that the order
parameter is conserved and the CH equation can be written in the 
form of a continuity equation, $\partial_tu=-{\bf \nabla.J}$, 
with the current ${\bf J}={\bf \nabla}(\delta f/ \delta u)$.
Therefore we may write the CH equation as following,
\begin{eqnarray}
\frac{\partial u}{\partial t} = -\nabla^2(\varepsilon^2\nabla^2u+u-u^3)
\label{EqCH}
\end{eqnarray}

Normally, $0<\varepsilon \ll 1$ because the fourth derivative
term requires a large stencil and it is related with the size
of the interface between regions of two different phase. To
accurately resolve these interfaces a fine space discretization 
$\Delta x$  is necessary. The linear
term is responsible for the interesting dynamics including the
instability of constant solutions near $u=0$ and the nonlinear
term is the one which mainly stabilizes the flow\cite{Eyre}. As it
has already been pointed out\cite{Eyre,Furihata,Elliot1,VLR},
both the $\nabla^4$ and the
nonlinear term make the CH equation very stiff and it is 
difficult to solve it numerically. The nonlinear term in principle, forbids the
use of common Fast Fourier Transform (FFT) methods and brings the 
additional problem that the usual
stability analysis like von Newmann criteria cannot be used. These difficulties
make most of the finite 
difference schemes to use time steps of many order of magnitude
smaller than $\Delta x$ and consequently, it is numerical expensive to reach
the time scales where the interesting dynamics occur. This is the
reason why numerical simulations based on Runge-Kutta schemes had to be performed in 
large supercomputers\cite{Toral0,Toral}.

In order to deal with such restrictions, Eyre developed a semi-implicit
method\cite{Eyre} which resolves the problems associated with both the 
stiffness and solvability. Furthermore Eyre proved that his algorithms
are unconditionally gradient stable for the CH and also for
the Allen-Cahn equation, which means that the free energy
does not increase with time. Both gradient stability and the
conservation of mass provide us with a simple and rational form to 
establish the  stability criterion for the CH equation that 
replaced the von Neumann stability criteria. Furthermore, mostly
either finite difference calculations\cite{Toral0,Toral} or
Monte Carlo simulations\cite{Binder} uses periodic boundary
conditions  as a tentative to mimic a large system and we
show here that, as concerns phase separation, that free boundary
conditions are less stiff and more faster achieved.

The goal of this paper is to make a combinations of a systematic 
study of the CH equation
in 1D, 2D and 3D using a simplification of the Eyre's method and
free boundary conditions. To
give a better perspective to  the approach, we make also calculations with a 
Crank-Nicholson like (CN) implicitly scheme\cite{Ames}, which is unconditionally
convergent in 1D and, using the concept of alternating direction interaction (ADI)
method\cite{Ames,Douglas}, in 2D. These calculations demonstrate the advantage of the 
Eyre's method over the CN scheme. Therefore we apply Eyre's method in
3D and study the phenomenon of spinodal decomposition  in three dimensions.
The application to the high $T_c$ superconductors and manganites with
the study of the relevant parameters to their phase diagrams is 
under current investigation and will be discussed in a future work.

\section{The Properties of the Cahn-Hilliard Equation}

In order to solve numerically Eq.(\ref{EqCH}) by means of a finite
difference scheme, we need an initial condition over the
entire domain, $u(\vec x,0)$, which usually is a random function or some
small fluctuations over a specific average and also some type of boundary conditions (BCs).
During our simulations we have found that these initial conditions, the
BCs and the size of the system greatly influence the
solution $u(\vec x,t)$.
The general convenient and flux-conserving boundary conditions are\cite{Furihata}
\begin{eqnarray}
{\bf \nabla}u.\vec n|_{\vec x\in\partial\Omega}=
(\nabla^3u).\vec n|_{\vec x\in\partial\Omega}=0 
\label{EqBC}
\end{eqnarray}
where $\vec n$ is the outward normal vector on the boundary of the domain
$\Omega$ which we represent by $\partial\Omega$. 
These two equations
together are equivalent to 
\begin{eqnarray}
{\bf\nabla} (\delta f/\delta u)_{\vec x\in\partial\Omega}=0 .
\label{EqBC2}
\end{eqnarray}
These BCs lead to the two very important properties
which will be our guide to know whether our  CH numerical solutions
are convergent, namely, the conservation of the total  mass of the
system $M_t$;
\begin{eqnarray}
\frac{d}{dt}M_t=&&\frac{d}{dt}\int_\Omega u(\vec x,t)d\vec x=\int_\Omega\frac
{\partial u(\vec x,t)}{\partial t} d\vec x 
\nonumber \\&&
=\int_\Omega \nabla^2\frac{\delta f}{\delta u}d\vec x=
\left[{\bf\nabla}\frac{\delta f}{\delta u}\right]_{\vec x\in\partial\Omega}=0
\label{EqCM}
\end{eqnarray}
and the dissipation or decrease of the total energy $F$;
\begin{eqnarray}
\frac{d}{dt}F(u)=&&\frac{d}{dt}\int_\Omega f(u(\vec x,t))d\vec x= \int_\Omega
\frac{\delta f}{\delta u}\frac{\partial u(\vec x,t)}{\partial t} d\vec x
\nonumber \\&&
=-\int_\Omega\left[\nabla\frac{\delta f}{\delta u}\right]^2d\vec x\le 0
\label{EqDE}
\end{eqnarray}

This last equation shows that appropriate solutions of the CH equation
must dissipate energy and this is called gradient flow.
Therefore a time stepping finite difference scheme is defined to be {\it gradient
stable} only if the free energy $F(u)$ does not increase with
the time, i.e., obeys Eq.(\ref{EqDE}). Since it is
not convenient to use the von Neumann stability analysis, gradient
stability is regarded as the best stability criterion for  
finite difference numerical solutions of the
CH equation \cite{VLR}. Furthermore, unconditional gradient stability
means that the conditions for gradient stability is satisfied
for any size of time step and this will be our guide through
the simulations. This way to examine the stability of a central
difference scheme by following the energy was already proposed
to nonlinear problems a long time ago by Park\cite{Park}.
We should mention that a similar analysis, based also on the Eyre's
approach, performing numerical tests of stability  and
with a very complete classification scheme for the stable values of $\Delta t$ 
for the CH and Allen-Cahn equation in 2D was recently developed\cite{VLR}.

\section{The Discretization Method}

Eyre has proposed a semi-implicit method that is unconditional
gradient stable if the  $V(u)$ is the usual two minima
potential used in a typical GL free energy and
can be divided in two parts:
$V(u)=V_c(u)+V_e(u)$ where $V_c$ is called contractive and 
$V_e$ is called expansive\cite{Eyre}. He showed that it is possible
to achieve unconditional gradient stability if one treats
the contractive part implicitly and the expansive part explicitly.
In our case, since we are using  $V(u)=(u^2-1)^2/4$, we have:
\begin{eqnarray}
 V_c(u)=\frac{u^4+1}{4} \hspace{0.5cm} and \hspace{0.5cm} V_e(u)=-\frac{u^2}{2} .
\nonumber
\label{VeVc}
\end{eqnarray}
To implement Eyre's scheme we define $U^n_{ijk} (i,j,k=1,2,...,N; n=0,1,2,...)$
to be the approximation to $u(\vec x,t)$ at location $x=ih$, $y=jh$, $z=kh$ and
$t=nK$, where $h=\Delta x=\Delta y=\Delta z$, $K=\Delta t$ and $N=L/\Delta x$. 
$L$ is the linear size of the system, assuming to be cubic,
for simplicity. With these definitions, we can
write the method for the CH equation  as
\begin{eqnarray}
\frac{U^{n+1}_{ijk}-U^n_{ijk}}{K}=-\varepsilon^2\nabla^4U^{n+1}_{ijk}+\nabla^2
\left((U^{n+1}_{ijk})^3-U^n_{ijk}\right).
\label{EqCHE}
\end{eqnarray} 
In this equation the standard centered difference approximation
of the 3D Laplacian operator $\nabla^2$ is 
\begin{eqnarray}
\nabla^2U^n_{ijk}=\left(U^n_{i+1jk}+U^n_{ij+1k}+U^n_{ijk+1}-6U^n_{ijk}+U^n_{i-1jk}
+U^n_{ij-1k}+U^n_{ijk-1}\right)/h^2
\label{EqLap}
\end{eqnarray} 
which is second order in the spatial step $h$.
The  Eq.(\ref{EqCHE}) represents a large coupled set of nonlinear equations
due to the cubic term. The way to go around this problem is to
splitting or to linearize it at every time step. Consequently the term 
$(U^{n+1}_{ijk})^3$ is transformed into $(U^{n}_{ijk})^2U^{n+1}_{ijk}$
leading the Eq.(\ref{EqCHE}) into a set of linear equations in the
step of time $n+1$. It has been
argued that this nonlinear splitting has the smallest local 
truncation error\cite{Eyre,VLR} and therefore it will be the
scheme adopted in this work. Then we finally  obtain the 
proposed finite difference scheme for the CH equation which
is linear (in the above sense), namely,
\begin{eqnarray}
\frac{U^{n+1}_{ijk}-U^n_{ijk}}{K}=-\varepsilon^2\nabla^4U^{n+1}_{ijk}+\nabla^2
\left((U^{n}_{ijk})^2U^{n+1}_{ijk}-U^n_{ijk}\right).
\label{EqCHEf}
\end{eqnarray}
or, separating in different times to see the semi-implicit character of the
approach,
\begin{eqnarray} 
U^{n+1}_{ijk}+K(\varepsilon^2\nabla^4U^{n+1}_{ijk}+\nabla^2(U^{n}_{ijk})^2U^{n+1}_{ijk})
= U^n_{ijk}-K\nabla^2U^n_{ijk}.
\label{EqCHEt}
\end{eqnarray}

As noted by Furihata et al\cite{Furihata}, the discrete  associated boundary conditions 
equivalent to Eq.(\ref{EqBC}), second order in the
spatial variable, become
\begin{eqnarray}
&& U^n_{1jk}=U^n_{5jk} \hspace{0.5cm}, \hspace{0.5cm} U^n_{2jk}=U^n_{4jk}
\\ && U^n_{Njk}=U^n_{N-4jk} \hspace{0.5cm}, \hspace{0.5cm} 
U^n_{N-1jk}=U^n_{N-3jk}
\label{EqUBC}
\end{eqnarray}
and similar equations for the boundary conditions over the second and
third indices representing the $y$ and $z$-directions, respectively.
Notice that, differently than most numerical simulations applied to 
physical systems\cite{Wang}, these boundary conditions are not
periodical. Imposing periodical BCs which is common in physical
applications and which is largely
used to artificially simulate a larger system, would bring
additional constraint to the solutions and, as we show below, the 
solutions will be more stiff. Therefore the above boundary conditions
minimize the finite size effects and should be preferably used
as shown below. 

The Eq.(\ref{EqCHEt}) with the above boundary conditions define the
finite difference scheme which we use in this present work. In the 
following section, we
analyze the numerical results and compare with  different
CN semi-implicit approaches for several dimensions with the
same double-well potential.

\section{The Results of the Simulations}

We start with the study of the CH equation in 1D because it is faster
then 2D and 3D and there are
many results which can be used to compare with our simulations.
The usual  semi-implicit and  explicit Euler's schemes for 
the 1D CH equation are not gradient stable and require very
short time intervals $\Delta t =K$, while the implicit Euler's
scheme is gradient stable with $K\le h^2/4=6.25 \times 10^{-6}$
\cite{Eyre,Elliot1,Mathews}. The CN-like scheme that we will use
for comparison also suffers
from this solvability restriction, and requires a 
minimum time interval  $K\le h^2/9$.

We performed calculations with linear chains  of $N=54, 104$ and $504$
sites and with $\Delta x=h=1/(N-4)$, and $0\le x \le1$. For all these cases we used
$K=h/2$ which is clearly several order of magnitude larger than 
typical Euler's schemes and, despite this large time step, the gradient
stability is observed at all times. In Fig.1 we show the results
for the total mass and the free energy $F$ in arbitrary units as 
function of the running  time. In fact, using
shorter times as shown in Fig.1, we observe that there is an initial 
transient period in the  time evolution before gradient stability is achieved. This
can be easily seen at at most small values of $\Delta t$ in Fig.1 for
either Eyre's or CN-like methods. 
This transient before the stability takes place is connected to 
the finite size of the system, the initial conditions and the BCs.

As already mentioned, in order to compare our calculations based on 
Eyre's method with other 
schemes, we have also performed similar
simulations with a widely used method\cite{Eyre,Elliot1},  
a semi implicit CN-like scheme\cite{Ames,Mathews}. Briefly, the CN method
consists in an average of the right hand side of Eq.(\ref{EqCH}) at
different times: $1/2$ at $n$ and $1/2$ at $n+1$ what clearly results in
a semi-implicit scheme in time.
According to the above formula, for $h=1/50$, this scheme will
converge for $K\le 1/(2500 \times9)\approx 1/20000)$. Indeed we can
see in Fig.1 that the CN simulations 
agree very well with the the Eyre's results for the
case with $K=1/50000$ and a similar transient period before
the stability is achieved is observed.
To compare with other studies which use the same parameters
for the CH equation, we used an initial condition for the linear chain similar
to one used by Furihata et al\cite{Furihata}, namely,
\begin{eqnarray}
  U^0_i=&&0.1sin(2.\pi(i-3)h)+0.01cos(4\pi(i-3)h)
\nonumber\\ &&+0.06sin(4\pi(i-3)h)+0.02cos(10\pi(i-3)h)
\label{EqU0}
\end{eqnarray}
for $3\le i \le N-2$ and $U^0_i=0.03$ otherwise (at the borders).
This function is shown in Fig.(\ref{FigUt1D}).

\begin{figure}[!ht]
\includegraphics[height=7cm]{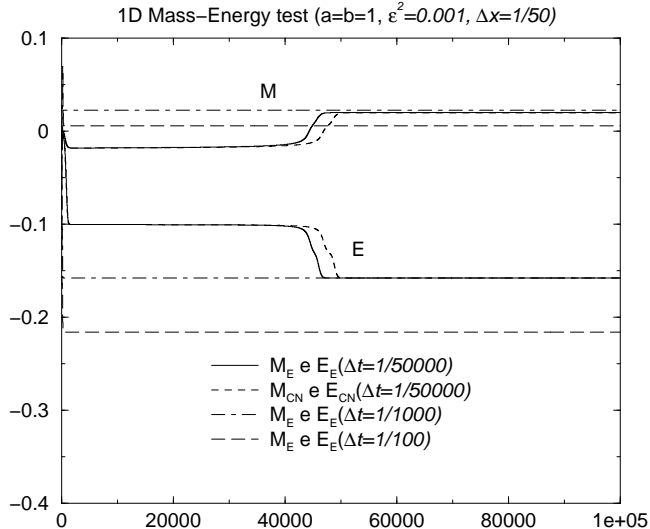}
\caption{ Time evolution of the total mass and the total energy 
in arbitrary units up to 
$10^5$ time steps (abscissa) by Eyre's and Crank-Nicholson's scheme
at different time steps for comparison.}
\label{FigME1D}
\end{figure}

Now that the parameters which gradient stability is established in
our system, we study the process
of spinodal decomposition in 1D starting with the initial condition
given by Eq.(\ref{EqU0}). Using the long time step $\Delta t=1/100$,
we see for the different $U^n_i$ time profile in Fig.2, that at very 
few time evolution ($n=5$) the system  shows a tendency to decompose
in one high and other low density phases. At $n=10$ the difference in
densities increases and the system separates in regions which almost reach the limit
$\pm1$ values. At $n=50$ and $n=100$ the system is almost
entirely separated in the  two $\pm1$ phases. At $n=500$ the limit configuration
is reached and the spinodal decomposition  is
total with the low density phase at left and the high phase at
right as seen in Fig.2. Furihata et al\cite{Furihata} have found a similar result 
starting from the same initial condition using a different method.
Larger systems behave in a similar fashion with the same sort of decomposition
seen in Fig.(\ref{FigUt1D}). Notice that the system separates in two regions
with different density (-1 at left near $x=0$ and +1 at the other and
at $x=1$). If we had  imposed periodic boundary
conditions, it would impose an additional constraint and the solutions
would be different than that two regions of Fig.(\ref{FigUt1D}), 
namely, three phase regions with one at the center and two of the
same kind at the borders. This more complex final configuration
takes more time steps to be realized as we checked, performing also calculations with
periodic boundary conditions. Thus, we
demonstrate that the used "free" boundary conditions are more "natural" and faster
than the periodic ones and this is {\it an important finding which will
be used in the 2D and 3D studies}.

\begin{figure}[!ht]
\includegraphics[height=7cm]{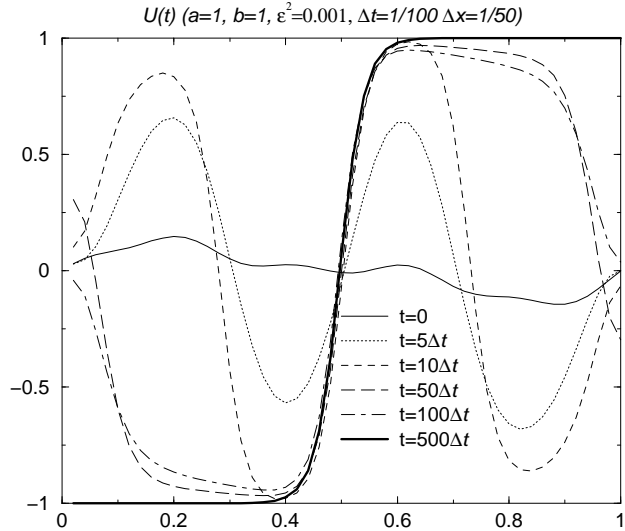}
  \caption{The time evolution of the profile of  $U^n_i$ for different 
times starting at the initial condition ($t=0$) till  the total spinodal 
decomposition is attained at $n=500$. The size of the system is $L=1$.}
\label{FigUt1D}
\end{figure}

Now, let's use what we learned above and turn our attention
to the 2D problem. We worked with systems of sizes
of $104\times104$ and $504\times504$ and again $\Delta x=h=1/(N-4)$
with $N=104$ or $504$. 
We adopt the Eyre's scheme
in the form of an alternating dimension implicit 
(ADI) problem\cite{Ames,Douglas}, that is, we
used a half marching time step in one direction, say, in the $x$-direction
and another half time step in the $y$-direction. Since most
finite difference methods for partial differential equations 
like the diffusion or heat equation in 2D uses the Crank-Nicholson
scheme\cite{Ames,Douglas}, we again made  simulations with
this scheme which is  appropriate to be used in connection
with the ADI\cite{Ames,Douglas,Elliot}.

In Fig.3 we plot the results for the 2D mass and energy in 
arbitrary units. We used $\varepsilon=0.01$ which is slightly bigger
than the 1D value in order to have larger phase domains and we kept
the other parameters equal. The initial conditions were chosen to
be small variation around the average value $U^n_{ij}=0.5$ although
variation around zero give the same type of result. Thus the
initial condition is
\begin{eqnarray}
U^0_{ij}=0.5+\varepsilon sin(\pi(i-3)20h)sin(\pi(j-3)20h)
\label{Eq2D0BC}
\end{eqnarray}
for $i,j$ toward the middle  and with $U^0_{ij}=0.5$
toward the boundaries of our square systems. The choice of a
different initial condition  than $U^0_{ij}=0$ is to brake
the symmetry and, in this way, better
identify the formation of the two $\pm 1$ density phases
(see Fig.(\ref{Fig2D2000}) below).
Studying the stability conditions, we see that the 2D system requires
smaller values of $\Delta t$ than the 1D system. 
This is because the boundaries are much larger in 2D and the 
possibility to mass and energy flow is greatly enhanced and the
instability during the initial transient period is larger.
For instance, the initial instabilities in the $504\times504$
is even bigger than in the $104\times104$ system.

\begin{figure}[!ht]
\includegraphics[height=6.5cm]{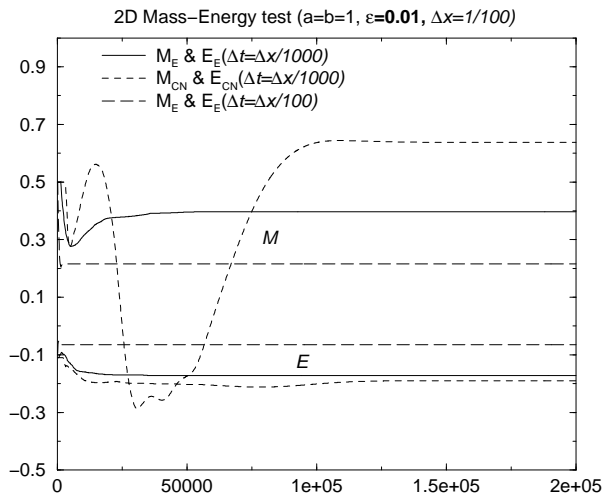}
\caption{ The stability of the mass and the energy in 2D for the 
$104\times104$ system by the Crank-Nicholson and Eyre's scheme 
for $\Delta x=1/100$ and  different time marching steps 
$\Delta t$ as specified in the legends.}
\label{Fig4}
\end{figure}

The CN scheme requires at least $\Delta t=\Delta x/1000$
which is around four times the minimum 1D value
and the Eyre's requires at least $\Delta t=\Delta x/50$.
In the simulations with the CN scheme, 
the mass oscillates wildly up to $10^5$ steps
and the energy has also a small increase near this number of
marching time steps. The Eyre's results for  $\Delta t=\Delta x/1000$ 
oscillate in the beginning transient of
the simulations but become gradient stable after 50000 steps
when the mass and the energy stabilize. On the other hand
the simulations are gradient stable and does not display any transient
oscillation since the beginning 
for $\Delta t=\Delta x/100$ but there is a large loss of average mass,
as shown in Fig.(\ref{Fig4}). 

As mentioned above the value of $\varepsilon$ must be higher than
that used in 1D in order to observe the phase separation phenomenon, 
otherwise the phase domains become very small. In order to study
the phase separation through the 2D Eyre's method (with 
$\Delta t=\Delta x/1000$), we started with
the initial condition of Eq.(\ref{Eq2D0BC}). As expected from the above
above analysis, around the $1000^{th}$ time step
the system start to separate into the $\pm 1$ density phases.
At the $1000^{th}$ time step the beginning of the spinodal 
separation is clear as shown in Fig.(\ref{Fig4}) below.
\begin{figure}[!ht]
\includegraphics[height=9cm]{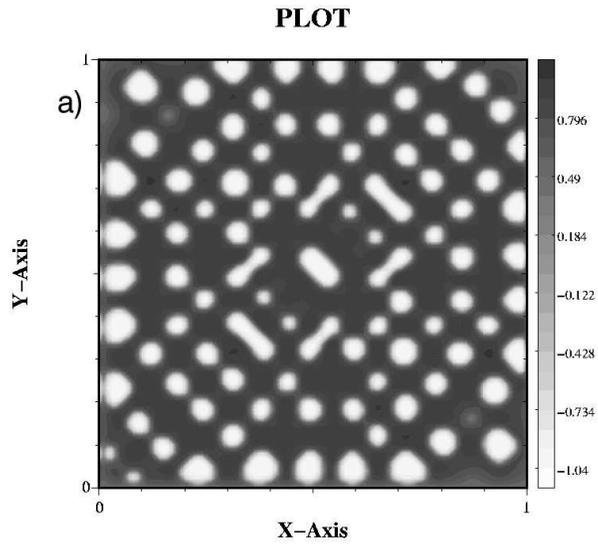}
\includegraphics[height=9cm]{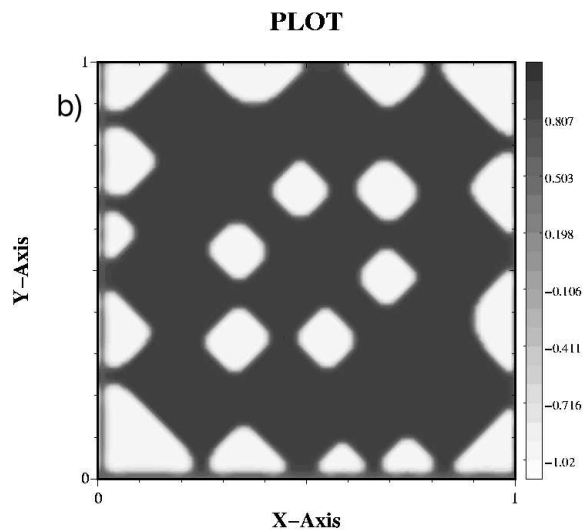}
\caption{ At the top panel, the beginning ($n=2000$) of the spinodal decomposition with the 
formation of islands with the values of -1 density in white in
a background which tends to +1 density (in black) because the
initial conditions with broken symmetry. At the panel below, we see the system at
a later time ($n=50000$) when it is already in equilibrium. The initial
configuration is that given by Eq.(\ref{Eq2D0BC}).}
\label{Fig2D2000}
\end{figure}

\begin{figure}
\includegraphics[height=9cm]{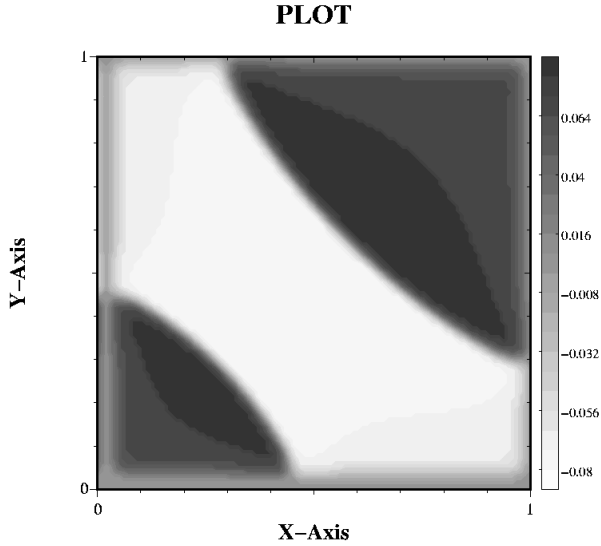}
\caption{ The same type of figure of Fig.(\ref{Fig2D2000})
but for $V(u)$ coefficients $a=0.6^2$ and $b=4.0^2$ (instead of $a=1$ and $b=1$, which produces
two easily separated phases with $u=\pm 0.15$) after
the equilibrium is achieved ($n=40000$).}
\label{Fig2Dlargeb}
\end{figure}

Around the $6000^{th}$ time step the  the size of the low density
islands reach an equilibrium size and the spinodal separation is
complete but, the boundaries effects are large. We see a concentration
of the low density phase at the borders and this effect prevents the
system to reach a total phase separation as we have seen in 1D (see
Fig.(2)). This shows how the boundary effects are very strong and
more significant in 2D.
After this time step the
system achieves a stable configuration without appreciable changes up
to $10^6$ time steps. Below we show a snap shot of this equilibrium
configuration at $n=50000$. Similar results were obtained with 
the $504\times504$ system but in this case there is a proportional
increase of the islands size and the equilibrium situation is similar
to five times blow up of Fig.(\ref{Fig2D2000}). Simulations with a larger value
of the non-linear coefficient $b$ (see discussion after Eq.(\ref{Eq2}))
enhances the mass flow inside the
system and it is possible to achieve complete phase separation at an
earlier time, exactly
as seen in 1D, but with two phases with smaller values than $\pm1$.
In Fig.(\ref{Fig2Dlargeb}) we plot a situation with $a=0.6$ and $b=4.0$.
In this case there is an easy  mass flow through the system and  an state of complete
phase separation is reached around $n=40000$. Notice that the system 
reaches the equilibrium with nonperiodical BCs.

We turn now to analyze the 3D system. As we have already found in the 2D
case, the problem of mass/energy flow is largely enhanced through the boundaries.
To deal with such effects, we have to use a small (compared with those used in the
2D case) marching time step of 
$\Delta t=\Delta x/10000$ with $\Delta x=1/100$ for the $(104)^3$ system 
to minimize errors in the derivatives
near the boundaries. Notice that this time step is much smaller than the
one in 1D and one-two orders of magnitude smaller than in 2D. This is small
but  still feasible to reach the interesting dynamics even 
in a typical PC of 1GHz. Since Eyre's method is very
efficient and faster than the majority of other methods used in 3D,
this is the only approach that we used in 3D.
The generalization of basic Euler or Runge-Kutta methods
to the CH equation in 3D are much more slower, and 
we can see why there are very few studies of the CH equation in
3D. Indeed the values of $\Delta t\approx 10^{-6}$ used in our 3D simulations are
of the same order of magnitude of typical Euler's method\cite{Eyre,Elliot1,Mathews}
used in 1D simulation as we discussed in the beginning of this section.
The adopted initial condition for the 3D system is:
\begin{eqnarray}
U^0_{ijk}=0.01+\varepsilon sin(\pi(i-3)20h)sin(\pi(j-3)20h)sin(\pi(k-3)20h).
\label{Eq3D0BC}
\end{eqnarray}
We used an average small initial mass of 0.01 just to minimize the mass
flow through the boundaries and
indeed the mass remains very stable for up to $4\times 10^4$ time steps,
when it starts a slightly increase. 
\begin{figure}[!ht]
\includegraphics[height=6.5cm]{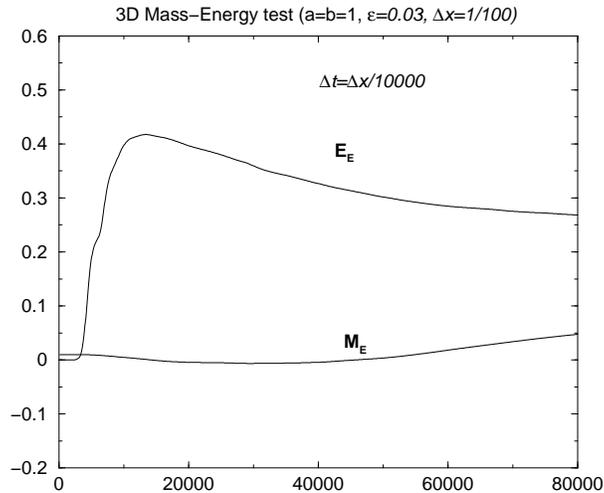}
\caption{ The stability of the mass and the energy in 3D for the
$104^3$ system by the  Eyre's scheme
for $\Delta x,y,z=1/100$.}
\label{Fig3Dmass}
\end{figure}

As concerns the energy,
it remains stable for the first step of the simulations but the
boundaries effects are manifested near the time step 4000 and increases up
to step 13000. During this transient period of time, although the mass is stable, the 
simulations are not, as one can conclude by the energy increase in this
interval as shown in Fig.(\ref{Fig3Dmass}).
After this time interval ($n \approx 13000$) the system stabilizes itself and
remains gradient stable up to the rest of the calculations (up to $n=80000$) 
as can be seen by the decrease of the energy in Fig.(\ref{Fig3Dmass}).
Comparison with the 1D and 2D systems reveals clearly that the stability is 
more difficult to be established as the dimensionality increases, as it
is natural when one uses any finite difference scheme.
\begin{figure}[!ht]
\includegraphics[height=9cm]{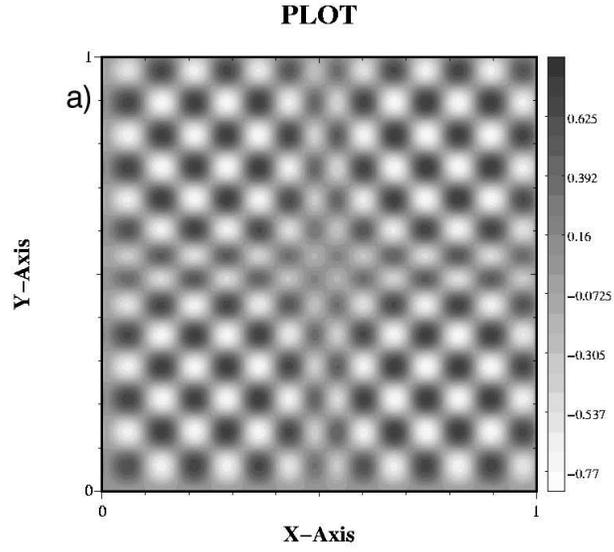}
\includegraphics[height=9cm]{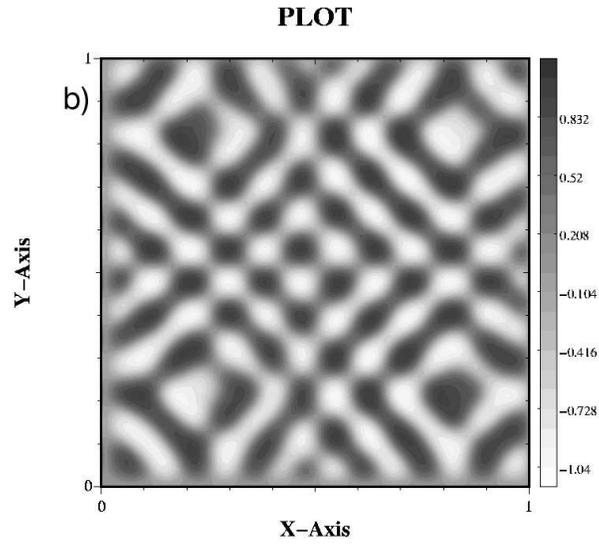}
\caption{ After relaxing from the initial condition (Eq.(\ref{Eq3D0BC})),
we show at the top panel the beginning ($n=6000$) of the spinodal decomposition with the
formation of small islands with the $\pm1$ values  because the
initial condition with broken symmetry. At the down panel, the system at
a later time ($n=8000$) during the process of phase separation.}
\label{Fig3Da}
\end{figure} 
As Fig.(\ref{Fig3Dmass}) shows, the system relax from the initial 
conditions and becomes uniform around $n=3000$. Around $n=6000$
a spinodal decomposition is happening and the oscillation in
the densities forms an almost uniform pattern as shown in 
the top panel of Fig.(\ref{Fig3Da}). This figure shows the configuration
of the middle plane in the z-direction ($k=52$) of the 
$104\times104\times104$. At the down panel of Fig.(\ref{Fig3Da})
we show a snap-shot of $n=8000$ which shows the beginning of 
of the  phase separation  process as the time flows. Above $n=20000$
the two phases start to segregate and this segregation process
is smooth and unconditionally gradient stable as seen in
Fig.(\ref{Fig3Db}). 

\begin{figure}[!ht]
\includegraphics[height=9cm]{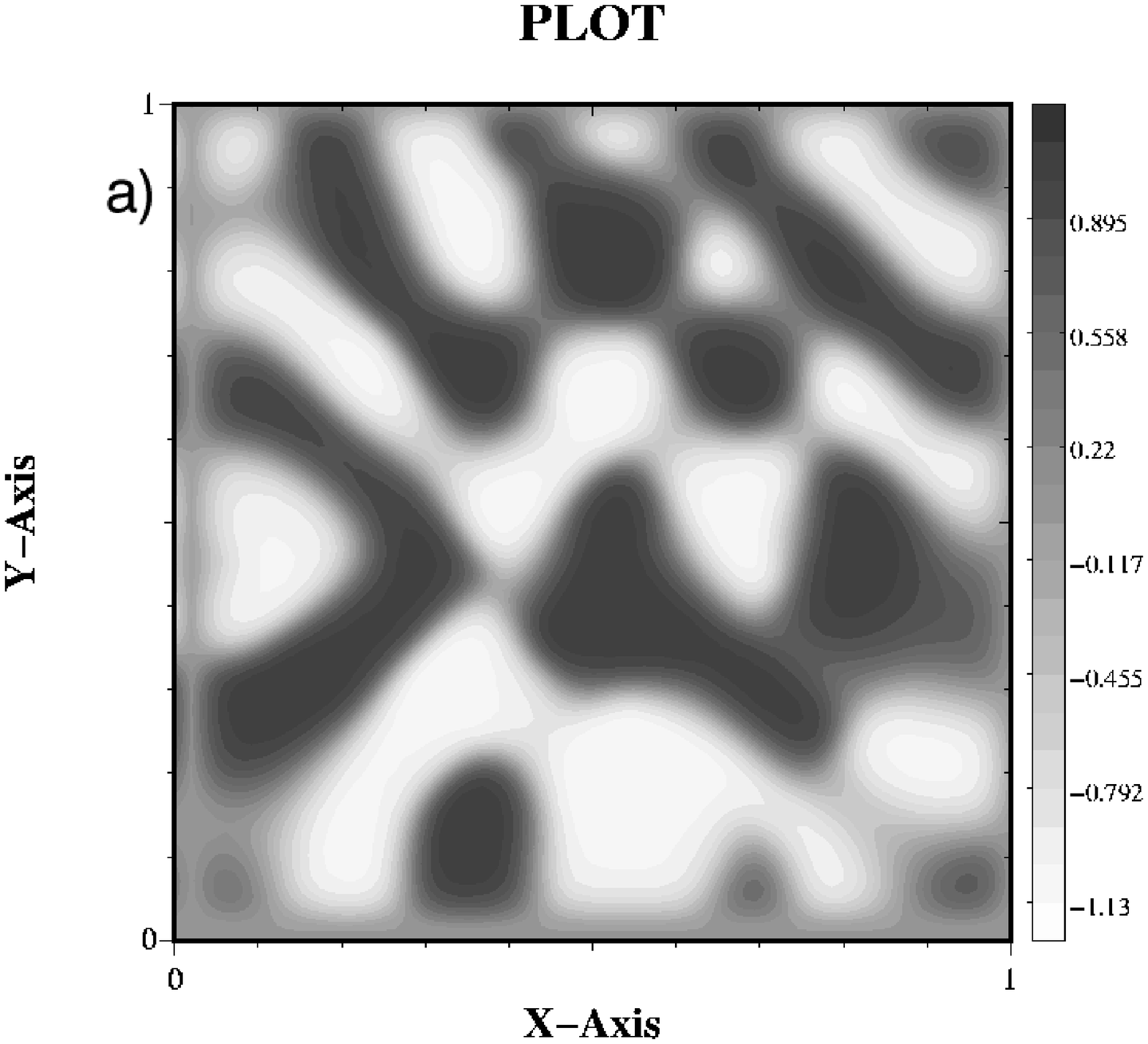}
\includegraphics[height=9cm]{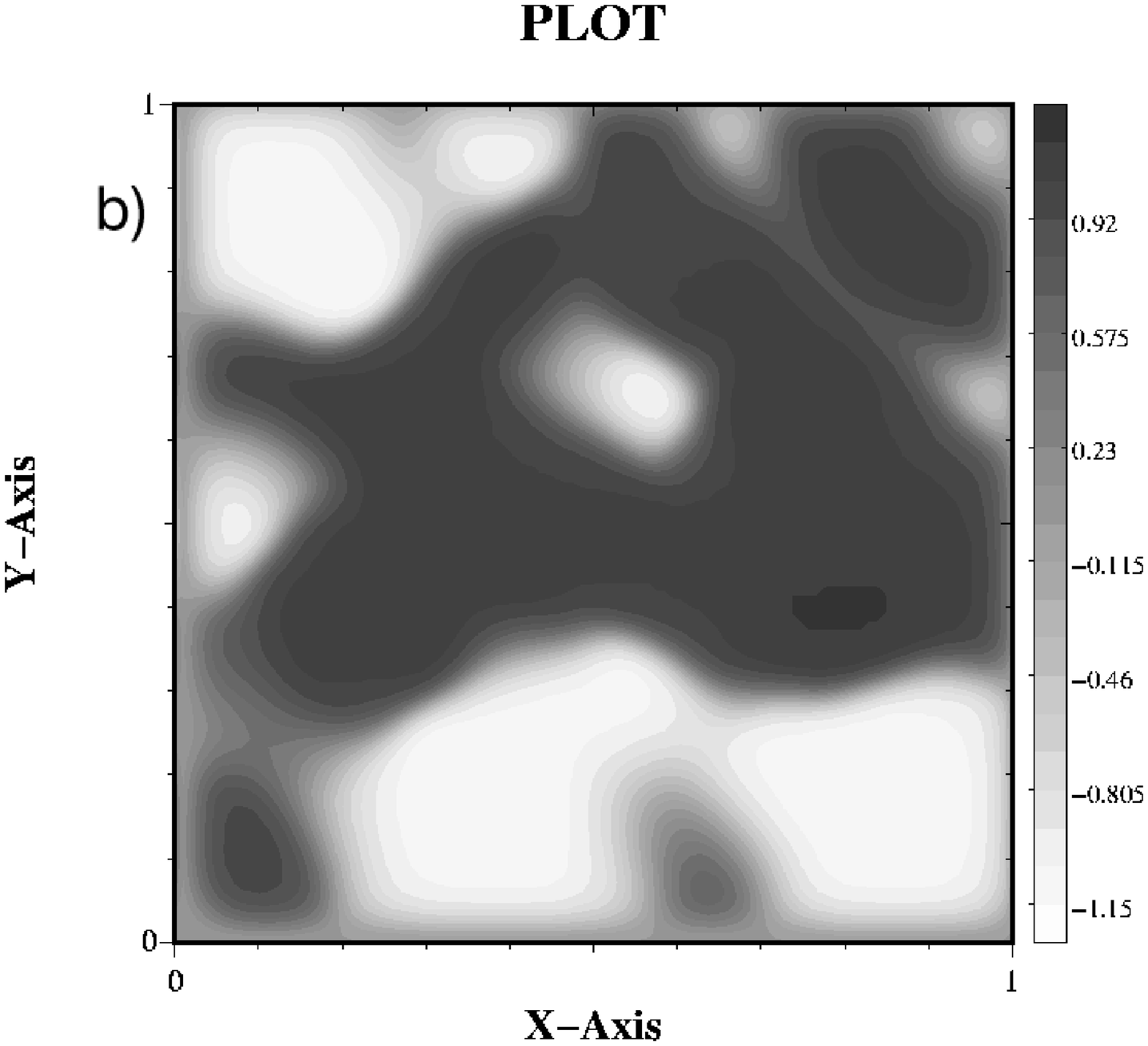}
\caption{ The continuation of Fig.(\ref{Fig3Da}) for many time steps later. On the
top panel we plot the case for $n=40000$ and on the down panel the $n=80000$.}
\label{Fig3Db}
\end{figure} 
At much later time like $n=40000$ the phase separation phenomenon is on
the way. At $n=80000$ the phase separation is almost total
in the plane passing by the middle of the system, as it is
shown in the down panel of Fig.(\ref{Fig3Db}).
\newpage
\section{Conclusion}

We have shown in this work that the semi-implicit method due to Eyre,
which divides the potential $V(u)$ in two parts and treats the
contractive part implicitly and the expansive explicitly combined
with free boundary conditions achieves
unconditional gradient stability in 1D, 2D and 3D. The method
also allows the use of very long marching time steps, compared
with usual explicit or implicit Euler's schemes, what is 
convenient because it captures very easily
the short and long dynamics, characteristic of the Cahn-Hilliard
equation. Our simulations have demonstrated that gradient stability
and spinodal decomposition is achieved faster than the normal Euler
and Crank-Nicholson methods what is highly deseirable specially in
3D where normal methods fail to capture the long dynamics due
to the required very short time steps.
The use of "nonflow"  free BCs are more appropriated and also
converges faster to a final equilibrium configuration than the widely used
periodic BCs.

We believe that the present systematic study may be pertinent
to the several branchs of physics. The fast scheme developed here
are suitable to the study of the spinodal dynamics and also to  
high correlated electron systems with phase separation. We expect 
therefore, to perform these
works in the near  future.

\section{Acknowledgment}
We gratefully acknowledge partial financial aid from Brazilian 
agencies CNPq and FAPERJ.

\end{document}